\documentclass[aps,twocolumn,showpacs,amsmath,amssymb,pra,floatfix,longbibliography]{revtex4-1}
\usepackage{ braket}
\usepackage{amsmath}
\usepackage{amssymb}
\usepackage{mathtools}
\usepackage{graphicx}
\usepackage{dcolumn}
\usepackage{bm}
\usepackage{multirow}
\usepackage{leftidx}
\usepackage{color}
\usepackage{qcircuit}
\usepackage{ifthen}

\usepackage{amsfonts}
\usepackage{eufrak}
\usepackage[german,english]{babel}

\newcommand \nn{\nonumber}

\newcommand \be{\begin{equation}}
\newcommand \ee{\end{equation}}
\newcommand \bea{\begin{eqnarray}}
\newcommand \eea{\end{eqnarray}}
\newcommand \bse{\begin{subequations}}
\newcommand \ese{\end{subequations}}

\newboolean{ShowComments}
\setboolean{ShowComments}{false}  

\definecolor{mscolor}{rgb}{0,0.5,0.5}

\begin{document}
\title{
Analysis of a Cesium lattice optical clock} 
\author{A. Sharma, S. Kolkowitz, and M. Saffman}
\affiliation{Department of Physics, University of Wisconsin-Madison,
Madison, Wisconsin 53706}

\date{\today}

\begin{abstract}
We propose and analyze a Cesium lattice optical clock (CLOC) which has the potential for high performance and simple operation in a compact form factor using a forbidden optical transition in Cs atoms at 685 nm. Cs atoms are trapped in a 3D optical lattice using a  magic trap wavelength of $\lambda_{\textrm m}=803~\textrm  nm$. To reduce sensitivity to magnetic fields the atoms are probed on two cycling transitions with equal magnitude, but opposite magnetic shifts. Operation of the clock requires only three diode lasers at 685, 803, 852 nm and is simplified compared to other higher-performance optical clocks that rely on alkaline earth atoms or single trapped ions. Analysis shows a quantum noise limited stability of $8.4\times 10^{-16}/\sqrt\tau$ and the potential for reaching 1 ns timing uncertainty at 1 month with realistic system parameters.
\end{abstract}

\maketitle

\section{Introduction}

Optical atomic clocks are the most precise timekeepers ever developed\cite{Ludlow2015}. These clocks have now reached fractional uncertainties roughly two orders of magnitude smaller\cite{Beloy2021} than the current SI time-keeping standard based on the microwave transition in the Cs atomic fountain clock\cite{Jefferts2007}.  High performance state-of-the-art optical atomic clocks remain complex and laboratory-scale devices that typically operate with low uptimes, in large part due to the complexity of the laser systems involved in cooling and repumping the atoms. For many well established and emerging applications of precision time keeping, reducing the size, weight, and power (SWaP) of the clock is critical. In this work we propose and analyze a cesium lattice optical clock (CLOC) based on a forbidden optical transition in Cs atoms. Our proposed clock therefore takes advantage of the relative simplicity of the level structure of alkali atoms and the well-established tool-box that has been developed to control them, while sacrificing some of the stability and accuracy offered by ultra-narrow forbidden transitions in more complex atoms and ions. We show that as a result, the proposed CLOC has the potential for performance exceeding current microwave standards in a low SWaP form factor and with simpler operation than current optical clocks.

Several recent reviews have surveyed progress in transportable high performance atomic clocks\cite{Delehaye2018,Gellesch2020}. Possible approaches include ion clocks \cite{JCao2017,Hannig2019,Burt2021}, microwave frequency cold atom clocks \cite{Farkas2010,Esnault2013,XLiu2017,Elvin2019}, and the optical frequency atomic clock based on a two-photon transition in Rb atoms\cite{Perrella2019,Maurice2020,Gerginov2018,Martin2018,Martin2019,Newman2019}. Of these approaches, our CLOC architecture is most similar to the two-photon Rb clock, which relies on lock-in detection of the fluorescence signal scattered from the upper level of the clock transition as opposed to Ramsey or Rabi interrogation. Recent demonstrations 
of this type of Rb clock have reached fractional uncertainties at integration time $\tau$ of $4\times 10^{-13}/\sqrt\tau$~\cite{Martin2018} and
$4.4\times 10^{-12}/\sqrt\tau$ in a compact form factor with photonic integration\cite{Newman2019}.

\begin{figure*}[!]
    \centering
    \includegraphics[width=0.8\textwidth]{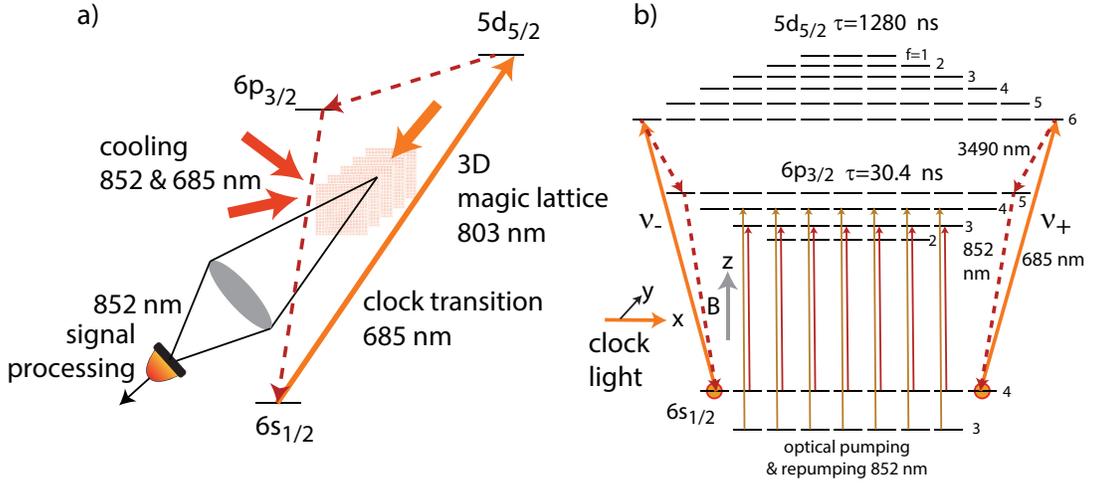}
    \caption{CLOC architecture. a)  The clock transition at 685 nm connects the $\ket{6s_{1/2},f=4}$ level to $\ket{5d_{5/2},f=6}$. The upper state decays with near 100\% branching ratio to $\ket{6p_{3/2},f=5}$   while emitting a 3490 nm photon, followed by decay back to $\ket{6s_{1/2},f=4}$ accompanied by emission of a 852 nm photon. Atoms are cooled at 852 and 685 nm and trapped in a 3D magic lattice with 803 nm light. The signal derived from detection of the 852 nm photons is used to stabilize the clock laser. b) Detailed atomic level diagram showing the participating states. The atomic quantization axis is defined by a bias magnetic field $B$.  Optical pumping and repumping light polarized along $z$ prepares the atoms in   $\ket{6s_{1/2},f=4,m=\pm4}$. The clock  probe light propagates along $x$ and is $y$ polarized. The probe drives the closed cycle $6s_{1/2}\ket{4,\pm4}\rightarrow5d_{5/2}\ket{6,\pm6}\rightarrow6p_{3/2}\ket{5,\pm5}\rightarrow6s_{1/2}\ket{4,\pm4}$. The $\pm$ transitions at frequencies $\nu_\pm=\nu_{\textrm c}\pm \nu_B$ are alternately sampled at the switching rate $f_{\textrm s}$. }  
    \label{fig.CLOC}
\end{figure*}

\section{Clock operation}

The proposed optical clock operates on the Cs $\ket{g}=\ket{6s_{1/2},f=4,m=\pm 4} \rightarrow \ket{e}=\ket{5d_{5/2},f=6,m=\pm 6}$ quadrupole (E2) transition near 685 nm, as shown in Fig.~\ref{fig.CLOC}.
As we describe below, this cold atom optical clock can be operated continuously, and requires only three laser wavelengths, 685, 803, and 852 nm, together with optical detection of emitted photons at 852 nm.

Atoms are cooled in a magneto-optical trap (MOT) with 852 nm light followed by second stage cooling using the clock laser on the clock transition which has a Doppler temperature of $3.0~\mu\textrm K$\cite{Carr2014t}. They are then loaded into a projected 3D optical lattice\cite{Huft2022} which is at a ``magic wavelength'' \cite{Takamoto2005} for the clock transition at a wavelength of $\lambda_{\textrm m}=803~\textrm  nm$. The 3D lattice only requires optical access from a single side which facilitates a compact form factor for the vacuum assembly.  The atoms are then pumped into the $6s_{1/2}\ket{4,\pm4}$ stretched states using 852 nm $x$ polarized light resonant with the $6s_{1/2},f=4 \rightarrow 6p_{3/2},f=4$ 
transition. The stretched states with $m=\pm4$ are dark states which minimizes heating during optical pumping. 

To drive the E2 transition a 685 nm clock laser propagates along $x$ and is linearly polarized along $y$.  This  configuration also couples $\ket{g}$ to $f=6,m=\pm2$ and $f=6,m=\pm4$ states but these transitions are strongly suppressed by Zeeman shifts from a bias magnetic field $B$.
Sensitivity to magnetic noise is suppressed by probing both the $\ket{4,4}\rightarrow \ket{6,6}$ and $\ket{4,-4}\rightarrow\ket{6,-6}$ transitions at frequencies $\nu_\pm=\nu_{\textrm c}\pm \nu_{B}$, where  $ \nu_{\textrm c}$ is the unshifted clock frequency and $\nu_B$ is the Zeeman shift from the bias field.  The sum $\nu_++\nu_-=2\nu_{\textrm c}$ is insensitive to magnetic noise, and the two transitions are alternately probed at a switching rate $f_{\textrm s}\sim 10~\textrm{kHz}$.

The excited states cycle back to the $6s_{1/2}\ket{4,\pm4}$ states emitting 3490 and 852 nm photons. The 852 nm photons are detected and used to feedback on the clock laser frequency and stabilize it to the atomic transition frequency $\nu_{\textrm c}.$ Selection rules ensure that the decay is cycling so that all atoms return to the $6s_{1/2}\ket{4,\pm4}$ states. To maintain a constant signal, atoms lost due to background collisions are periodically replenished by loading from the MOT. By interleaving loading with alternate probing of two samples continuous operation of the clock can be maintained  \cite{Biedermann2013,Schioppo2017}. 

\subsection{Clock stability analysis}
\label{sec.stability}

The CLOC architecture uses detection of the scattered 852 nm signal for stabilization of the 685 nm laser to the quadrupole transition. The clock uncertainty is fundamentally limited by detector shot noise resulting in a stability in fractional frequency units of 
\be
\sigma(\tau)= \frac{\Delta\nu}{\nu_{\textrm c}}\sqrt{\frac{1}{\dot N\tau}}.
\label{eq.sigma}
\ee
Here $\Delta\nu$ is the full width at half maximum (FWHM) transition linewidth,
$\nu_{\textrm c}$ is the clock  frequency, $\dot N$ is the number of detection events per second, and $\tau$ is the integration time. 

To estimate the shot noise limited performance we use $\nu_{\textrm c}=4.376\times 10^{14}~\textrm{Hz}$ for the clock transition frequency. At unit saturation $\Delta\nu=\sqrt2/(2\pi \tau_{\textrm a})=1.75\times 10^5~\textrm{Hz}$, with $\tau_{\textrm a}=1.28~\mu\textrm s$ the $5d_{5/2}$ radiative lifetime. The detection rate is 
\begin{equation}
\dot N= \eta_{\textrm col}\eta_{\textrm det}\frac{1}{4\tau_{\textrm a}}N_{\textrm a}
\label{eq.Ndot}
\end{equation} 
where $N_{\textrm a}$ is the number of atoms, the optical  efficiency is $\eta_{\textrm col}=0.2$ corresponding to collection optics with numerical aperture NA$=0.8$,  the detector quantum efficiency is $\eta_{\textrm det}=0.9$ for a Si detector at 852 nm, and $1/4\tau_{\textrm a}=1.95\times 10^5~\textrm s^{-1}$ is the scattering rate per atom  at unit saturation.  

With $N_{\textrm a}=6.6\times 10^6$ atoms (see Sec.~\ref{sec.traparray} for details of the trap design) and $\dot N=2.3\times 10^{11}~\textrm s^{-1}$ eq.  (\ref{eq.sigma})  evaluates to  
\be 
\sigma(\tau)=\frac{8.4\times 10^{-16}}{\sqrt\tau}.
\label{eq.sigmaresult}
\ee
This is the standard quantum limit for the clock stability. 
The corresponding frequency uncertainty at 1 s is $\sigma \nu_{\textrm c}=0.37~\textrm{Hz}$.

In addition to the quantum limit the short term stability is impacted by the intermodulation effect\cite{Audoin1991} at frequencies of $2f_{\textrm s}$ corresponding to switching between the $\nu_\pm$ transitions and $2f_{\textrm m}$ corresponding to the modulation frequency used for lock in detection of the photodetector current. We anticipate using $f_{\textrm s}=10~\textrm{kHz}$ and $f_{\textrm m}=100~\textrm{kHz}$. The intermodulation frequency contribution to the frequency deviation is
\begin{equation}
\sigma^{(\textrm IM)}    =\frac{\left[S(2f)\right]^{1/2}}{2\nu_c\sqrt\tau}
\label{eq.sigmaIM}
\end{equation}
with $S(2f)$ the power spectral density of  frequency noise at  $2f$ in units of $\textrm{Hz}^2/Hz$.
There are two primary contributions 
to $S_y$: the frequency noise of the probe laser and the shot noise of the detector signal. For the probe laser we  assume that at frequencies above 20 kHz there is white frequency noise of $1~\textrm{Hz}^2/Hz$. This level of frequency noise has been achieved with optically injected slave laser diodes\cite{YYLiu2021}. The correspondsing  local oscillator  contribution to the stability is
\begin{equation}
  \sigma^{(\textrm IM, LO)} = 1.1\times 10^{-15}/\sqrt\tau.
  \label{eq.sigmaIM1}
\end{equation}

The detector photocurrent is $I=e \dot N= 37~\textrm nA$. This implies a shot noise limited signal to noise ratio of SNR$=3.4\times 10^5~\textrm (Hz)^{-1/2}$. We estimate the corresponding  frequency deviation as
\begin{equation}
\sigma^{(\textrm IM, shot)} 
=\frac{\left[\Delta\nu/{\textrm SNR}\right]^{1/2}}{2\nu_c\sqrt\tau}
= 8.2\times 10^{-16}/\sqrt\tau.
  \label{eq.sigmaIM2}
\end{equation}
Adding eqs. (\ref{eq.sigmaresult}, \ref{eq.sigmaIM1}, \ref{eq.sigmaIM2}) in quadrature gives a short term stability limit of $\sigma=1.6\times 10^{-15}/\sqrt\tau$.  
The achievable stability at long times
depends on the level of control over systematic shifts, and is discussed in Sec.~\ref{sec.systematics}.

\section{Experimental Design}

\begin{figure}[!t]
    \centering
    \includegraphics[width=0.45\textwidth]{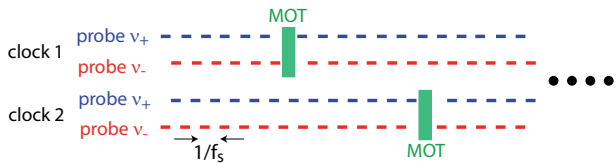}
    \caption{Continuous clock operation using two cells that are alternately probed at frequencies $\nu_\pm$. A periodic  MOT and optical pumping phase  replenishises lattice atoms. }
    \label{fig.sequence}
\end{figure}

We envision a clock based on alternating interrogation between two separate vacuum cells, or time alternating interrogation of a single atomic sample. The timing sequence is shown in Fig.~\ref{fig.sequence}.  A pre-cooled, and occupied atomic lattice will be prepared in each cell. As discussed in the following the lattice will be addressed with alternating probing and cooling cycles. Using two cells and out of phase probing and cooling a continuous clock signal will be available. The Zeeman shifted $\pm$ transitions are alternately probed at a switching rate of $f_{\textrm s}$. Each transition is detected by modulating the probe light at frequency $f_{\textrm m}\gg f_{\textrm s}$ followed by lock in detection of the scattered 852 nm light to provide an error signal for the laser frequency.  On a time scale of a few seconds one of the lattices will be refreshed from a  MOT followed by second stage cooling and optical pumping.  In this way a continuous clock measurement can be maintained indefinitely. The continuous and high bandwidth photon signal relaxes the requirements on the clock laser, potentially obviating the need for a ultra-high-stability bulky and vibration sensitive cavity.

\subsection{Laser Cooling}

We propose a two-step cooling process starting with a standard magneto-optical trap (MOT) at 852 nm driving the $6s_{1/2},f=4\rightarrow 6p_{3/2},f=5$ transition. This can be performed with a single diode laser by adding a 8.94 GHz sideband for repumping from $f=3$. Cold clouds of $> 10^6$ atoms can be prepared in $\ll 1~\textrm s$ using this approach. 

Second stage cooling uses  frequency shifted clock laser light tuned to the red of the 
$6s_{1/2},f=4\rightarrow 5d_{5/2},f=6$ transition which has a Doppler temperature of $3.0~\mu\textrm K$. Preliminary demonstrations of cooling on this transition as well as analysis of sub-Doppler cooling can be found in \cite{Carr2014t}. Decay from the excited state proceeds via the closed cascade $ 5d_{5/2},f=6 \rightarrow 6p_{3/2},f=5\rightarrow 6s_{1/2},f=4$ with essentially 100\% branching ratio. The depumping fraction is $\sim 10^{-7}$ (see Sec.~\ref{sec.magic}) so a repumper is not needed. We conservatively predict cooling to $T\sim 1~ \mu\textrm K$ which corresponds to about $10$ times the recoil temperature. 

\subsection{Trap design}
\label{sec.traparray}

We propose to use a three dimensional array of dark optical traps that confine atoms near zeros of the optical intensity. Such arrays are traditionally prepared using six counterpropagating beams. The implementation can be simplified using a 
a recently demonstrated projected ``hole" array\cite{Huft2021}. The array is prepared using a partially transmitting optical mask followed by  $4f$ Fourier filtering to create a two-dimensional pattern of near Gaussian holes in a uniform background. The magnification of the optics used to project the light into the vacuum cell is chosen to give an array period of $d$. The Talbot effect leads to repetition of the planar array at axial spacings of the Talbot length 
$L_{\textrm T}=\lambda/[1-\sqrt{1-\lambda^2/d^2}]$, where $\lambda$ is the wavelength of the trapping light. Taking $d=0.9~\mu\textrm m$ and $\lambda=0.803~\mu\textrm m$ gives $L_{\textrm T}=1.46~\mu\textrm m$. 
We assume a cube of size  $w=250~\mu\textrm m$ on a side. This gives a total of $w^3/(d^2 L_{\textrm T})=1.3\times 10^7$ trapping sites. Assuming a single atom filling fraction of 0.5 we obtain $N_{\textrm a}=6.6\times 10^6$ atoms. In this regime broadening due to atomic collisions is strongly suppressed. 
  
At the magic wavelength of $\lambda_{\textrm m}=803 \textrm{ nm}$ the $6s_{1/2}$ ground state has a polarizability of $\alpha_0=-374.~\textrm \AA^3$. With the trap region size of $250~\mu\textrm m$ on a side an optical power of $P=2~\textrm W$ gives a trap depth of $18~\mu\textrm K$ which is about 145 times the recoil energy. This power level is available from semiconductor tapered amplifier devices which provide power up to 3 W at 800 nm.  In order to keep the optical power requirement in the range of 
readily available tapered amplifier components we propose a compact lattice and a  relatively high atomic density. This in turn implies a large optical density which will inhibit full optical pumping and also lead to depumping from reabsorption of scattered 852 nm photons during clock operation. Cycling photons emitted during the $6p_{3/2}\ket{5,5}\rightarrow 6s_{1/2}\ket{4,4}$ decay are reabsorbed on the same  
transition with cross section $\sigma_0=3\lambda_{852}^2/(2\pi).$ Since the emitted photons are distributed over the full solid angle they can be reabsorbed by driving $\pi$ and $\sigma_-$ transitions to $6p_{3/2}\ket{5,4}$ and $\ket{5,3}.$ The absorption cross section for these transitions is $\sigma_0 \times 1/5$ and $\sigma_0 \times 1/45$. The unwanted transitions are further suppressed by Zeeman shifts relative to $\ket{5,5}.$ Assuming a bias magnetic field of $1$~mT We estimate the effective optical depth of the medium for a
photon emitted at the center of the lattice as 6.5 with the
optical depth  due to depolarizing $\pi$ and $\sigma_-$ transitions only 1.3. The depolarizing rate can in principle be suppressed further at the cost of higher optical power to support a more dilute optical ensemble. We defer a more detailed analysis of the impact of photon reabsorption on clock performance to future work.

\begin{figure}[!t]
    \centering
    \includegraphics[width=0.4\textwidth]{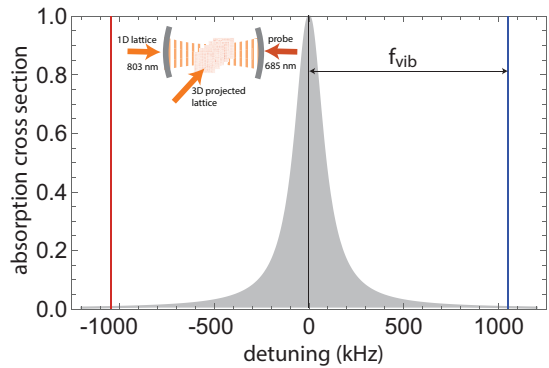}
    \caption{Absorption profile of clock transition at unit saturation. Red and blue vibrational sidebands are shown for the parameters given in the text. }  
    \label{fig.sidebands}
\end{figure}

In order to suppress Doppler broadening and recoil heating the clock should be operated in the resolved sideband limit with a small Lamb-Dicke parameter $\eta$.  The excited state linewidth is $\Delta\nu/2\pi=124~\textrm{kHz}$ which we require to be small compared to the trap vibrational frequency along the axis of the probe beam. To achieve a high vibrational frequency we propose to use an additional standing wave 1D lattice with cavity build up, as shown in Fig.~\ref{fig.sidebands}. With a 1D lattice beam power of 2.2 W, cross sectional area of $w^2$, and modest cavity build up factor of 50 we obtain an axial vibrational frequency of 1.05 MHz. When the clock transition is probed at unit saturation the relative absorption at the first vibrational sideband is a negligible 0.007.  The Lamb-Dicke parameter is 0.078 which adds further suppression to the sideband excitation. 

The projected lattice and 1D lattice beams propagate in orthogonal directions with the same electric field polarization. The beams can be detuned by a few MHz to remove interference effects, without compromising the long time clock stability (see Sec.~\ref{sec.latwavestab}).

\subsection{Quadrupole clock transition}

The quadrupole clock transition will be probed with 685 nm light that propagates along $x$ and is $y$ polarized. With this geometry transitions $m\rightarrow m'=m-2,m+2$ are allowed.  The matrix element for the transitions $\ket{4,\pm4}\rightarrow \ket{6,\pm6}$ is larger than that for the transitions $\ket{4,\pm4}\rightarrow \ket{6,\pm2}$ by a factor of $3\sqrt{55}=22.2.$ Thus the transitions between stretched states are strongly favored and  the saturation intensity is $570.~ \textrm mW/cm^2$.  The unwanted transitions are further suppressed by tensor shifts in the $5d_{5/2}$ level and by Zeeman shifts.   
For a beam of cross sectional area $w^2$ matching the size of the atomic lattice this corresponds to a modest probe power of 0.36 mW.

Since the clock transition is cycling the depumping rate into the ground $f=3$ level is dominated by scattering from $5d_{5/2},f=5$ at an approximate  rate of 
$$
\gamma_{\textrm dp}=\frac{\gamma}{2} \frac{1}{2+4\Delta_{65}^2/\gamma^2},
$$
where we have assumed unit saturation of the clock transition and the excited state hyperfine splitting between $f=5$ and $6$ is $\Delta_{65}=2\pi\times 127~\textrm{MHz}$. We find $\gamma_{\textrm dp}/\gamma = 1.2\times 10^{-7}$ which implies a depumping rate per atom of $0.004~\textrm s^{-1}$ or a depumping time of $270~\textrm s.$ This is significantly longer than the expected vacuum lifetime in current generation small form factor UHV cells, so the cycle time for replenishing the atomic lattice will be governed by vacuum quality, not depumping.

\subsection{Magic optical trapping}
\label{sec.magic}

The dynamic Stark shifts of the clock states can be readily calculated using a standard sum over states formalism accounting for all electric dipole allowed transitions. 
Matrix elements and transition frequencies were taken from \cite{UDportal,Sansonetti2009}. A core correction to the scalar polarizabilities was included using the value of $\alpha_{\textrm c}=15.84~ a_0^3$ from \cite{Safronova2016}, and $a_0$ is the Bohr radius.
The hyperfine resolved polarizabilities of the $5d_{5/2}$ states were calculated using 
\begin{eqnarray}
\alpha(f,m)&=&\alpha_0^{(j)}+\frac{3m_f^2-f(f+1)}{f(2f-1)}\alpha_2^{(f)}\\
 \alpha_2^{(f)}&=& \alpha_2^{(j)} \frac{3K(K-1)-4f(f+1)j(j+1)}{(2f+3)(2f+2)j(2j-1)}
 \end{eqnarray}
where $\alpha_0^{(j)}, \alpha_2^{(j)}$ are calculated in the fine structure basis as outlined above and 
 $K=f(f+1)+j(j+1)-I(I+1)$, with $I=7/2$  the nuclear spin. 

\begin{figure}[!t]
    \centering
    \includegraphics[width=0.5\textwidth]{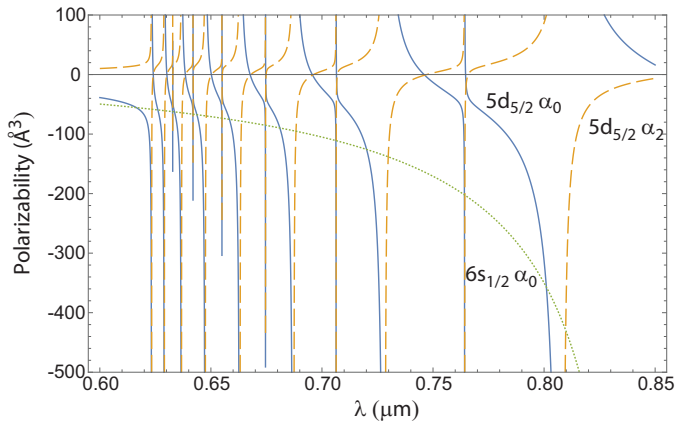}
    \caption{Polarizability of ground $6s_{1/2}$ and excited $5d_{5/2}$ states.
    The solid blue, dashed yellow, and dotted green lines show the $5d_{5/2} \alpha_0^{(j)}$,  $5d_{5/2} \alpha_2^{(j)}$, and $ 6s_{1/2} \alpha_0^{(j)}$ polarizabilities respectively. }
    \label{fig.polarizability}
\end{figure}

\begin{figure}[!t]
    \centering
    \includegraphics[width=0.45\textwidth]{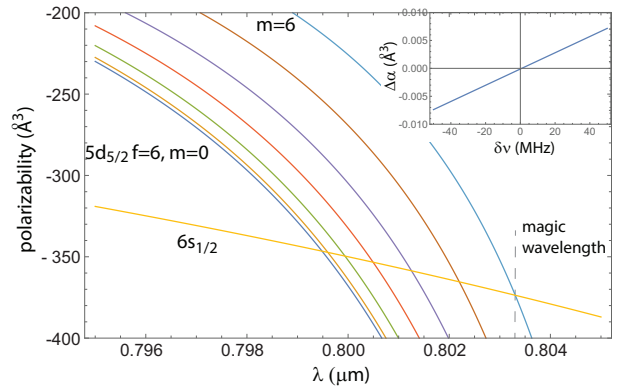}
    \caption{Polarizability of ground and excited states near $0.8~\mu\textrm m$. The dashed line shows the magic wavelength at $\lambda_m=0.8033~\mu\textrm m$ for the transition $6s_{1/2} \rightarrow 5d_{5/2},f=6,m_f=6$. The inset shows the differential polarizability as a function of detuning from the magic point.   }
    \label{fig.dynamicStark2}
\end{figure}

Figure \ref{fig.polarizability} shows that there are a large number of magic  wavelengths for trapping the $6s_{1/2}, 5d_{5/2}$ states between 0.6 and 0.9 $\mu\textrm m$ at which the states have a negative polarizability suitable for  dark optical traps. Magic conditions occur for wavelengths near where $\alpha_0$ is the same for the ground and excited state. In order to minimize the optical power needed for trapping we choose the magic condition with the largest polarizability near $0.8~\mu\textrm m$. The Zeeman state resolved polarizabilities for the $5d_{5/2}, f=6$ level are shown in Fig.~\ref{fig.dynamicStark2}.  The resonance behavior of the $5d_{5/2}$ polarizability is due to the transition to the $5f$ levels at 808 nm.

The tensor polarizability of the $5d_{5/2}$ level results in differential shifts between the Zeeman substates of the $f=6$ level. To estimate the severity of these shifts we assume that the atoms are cooled to $T=1~\mu\textrm K$, which is $3\times$ lower than the $3~\mu\textrm K$ Doppler temperature of the $6s_{1/2}-5d_{5/2}$ cooling transition. Calculations have confirmed the presence of sub-Doppler cooling on this transition\cite{Carr2014t}. The fractional spread of the polarizability at the magic wavelength for the $m_f=0$  excited state is $\eta=|[\alpha(6,6)-\alpha(6,0)]/\alpha(6,0)|=0.39$. This implies that an atom in a dark trap sees a spread of transition frequencies of $\eta \frac{k_B T}{2\pi \hbar}= 8.1~\textrm{kHz}$. This is a factor of 15 smaller than the excited state linewidth and has a negligible impact on the short term clock stability. However, the tensor shifts result in a strong sensitivity to trap light intensity that impacts long term systematics. For this reason we will only use transitions between the stretched states $\ket{4,\pm 4}\rightarrow \ket{6,\pm 6}$ which have a magic condition at 803 nm as shown in Fig.~\ref{fig.dynamicStark2}.

\subsection{Magic magnetic field}
\label{sec.magnetic}

Magic magnetic field conditions can  be 
found from a Breit-Rabi type analysis applied to the $6s_{1/2}$ and $5d_{5/2}$ levels. The $6s_{1/2}, f=4$
level has $g_f=1/4$ and  $5d_{5/2}, f=6$
 has $g_f=1/2.$ We therefore expect magic conditions to occur at low fields for transitions from $f=4$ to $f'=6$ for Zeeman substates 
\bea
m_f=\pm 2 \rightarrow m_f'=\pm 1,\nn\\
m_f=\pm4 \rightarrow m_f'=\pm2\nn.
\eea
There are also magic conditions at higher field values that rely on substantial mixing of the $f'=5,6$ levels. 

We have identified multiple magic magnetic field values for a variety of pairs of ground and excited Zeeman states. However, while it is possible to selectively tune the probe laser to a magnetic field dependent pair of states at a magic field condition, the atomic decay back to the ground state is not cycling with respect to the ground Zeeman substate which would lead to rapid depumping of the probed transition. Repumping of the clock states is an option, but this would result in additional dead time and increased complexity.

We therefore  propose to instead probe the $6s_{1/2}\ket{4,\pm4}\rightarrow 5d_{5/2}\ket{6,\pm6}$ transitions with a beam that propagates along $x$ and is polarized along $y$. This geometry couples the ground states  $\ket{4,\pm4}$ to the excited states $\ket{6,\pm2},  \ket{6,\pm6}.$
Using a moderate magnetic field we can ensure that only the $\ket{6,\pm6}$ states are excited which decay in a cycling fashion back to $\ket{4,\pm4}$. With a  magnetic field oriented along $z$ these states do not suffer Zeeman mixing so the sensitivity of the clock transitions to magnetic noise is
$$
\delta\nu_\pm(m)=\pm ( 6 g_{5d_{5/2},f=6}-4 g_{6s_{1/2},f=4})\frac{\mu_B}{2\pi\hbar} \delta B
=\pm\frac{\mu_B}{\pi\hbar} \delta B
$$
with $\mu_B$ the Bohr magneton and $\delta B$ the magnetic field fluctuation. A moderate magnetic field stability of $\delta B=10^{-8}~\textrm T$  results in 
$\delta \nu_\pm=\pm 280~\textrm{Hz}$. This is much less than the excited state linewidth and gives a negligible contribution to the short term clock stability, but is a limiting factor for long term stability. To circumvent this issue we will probe both stretched states and use the average frequency which is insensitive to the magnetic field strength.

\begin{table*}
    \centering
    \caption{Systematic uncertainties affecting the clock transition. The right hand column gives the stability requirement for 1 ns timing uncertainty at 30 days. The requirement on magnetic field variation corresponds to probing a single transition on  two separate clock ensembles. Alternating probing of $\nu_\pm$ on a single or two ensembles removes this requirement.    }
    \begin{tabular}{|l|l|c|} 
    \hline
shift & fractional coefficient $\beta/\nu_{\textrm c}$& 1 ns stability at 30 days\\ 
    \hline
probe beam power & $2.4\times 10^{-16}~/\%$ & 1.6 \%\\
lattice laser frequency shift from magic value &$ 1.8\times 10^{-17}/\textrm{MHz} $  &  21.6 MHz    \\  
magnetic field spatial variation & $3.2\times 10^{-12}~\textrm/(10^{-7}~\textrm T)$  & $12~\textrm{pT}$     \\  
dc electric field &$-1.1\times 10^{-14}~/\textrm{(V/m)}$ & $0.036~\textrm{V/m}$ \\
blackbody radiation &$2.7\times 10^{-16}~/\textrm K$ & $1.4~\textrm K$\\
    \hline
    \end{tabular}
    \label{tab.parameters}
\end{table*}

\section{Systematic uncertainties}
\label{sec.systematics}

The clock frequency is sensitive to parameters $a_i$ according to  $\delta\nu = \sum_i \beta_i \delta a_i$
where $\beta_i=d\nu_{\textrm c}/da_i$ and  $a_i$ are environmental  parameters. The  required stability of the $a_i$ to ensure a timing uncertainty of $\delta t$ after integration time $\tau$ is found from 
$$
\delta t = \frac{\delta\nu}{\nu_{\textrm c}}\tau
 =\frac{\sum_i \beta_i\delta a_i}{\nu_{\textrm c}}\tau
$$
so 
$$
\delta a_i=\frac{\nu_{\textrm c}}{\beta_i \tau}\delta t_i
$$
and $\delta t=\sum_i \delta t_i$.
Table \ref{tab.parameters} lists sources of systematic uncertainty, and the required stability for each parameter to reach a timing uncertainty of 1 ns at 30 days, corresponding to a long-term precision and accuracy of $3.9\times10^{-16}$. The fractional coefficients listed in Table \ref{tab.parameters} are $\beta_i/\nu_{\textrm c}$. Precise, but not unrealistic, control of environmental parameters will be needed to reach this level of performance.

\subsection{Magnetic field}
\label{sec.Bfieldsystematic}

As described above the clock transition will be probed on $6s_{1/2}\ket{4,4}\rightarrow 5d_{5/2}\ket{6,6}$
and $6s_{1/2}\ket{4,-4}\rightarrow 5d_{5/2}\ket{6,-6}$. With a bias magnetic field $B$ along the $z$ axis these states have a purely linear Zeeman shift. The differential shifts of the $\ket{4,\pm 4}\rightarrow\ket{6,\pm 6}$ transitions are  $$\delta\nu_\pm=  \pm \frac{\mu_B}{\pi\hbar} B.
$$
 At fixed magnetic field the average of the shifts is zero, which can be used to determine the unshifted clock frequency. 

We envision two different approaches to probing both transitions. Applying $z$ polarized pumping light on the $6s_{1/2},f=4 \rightarrow 6p_{3/2},f=3$ transition, together with a repump sideband driving $6s_{1/2},f=3 \rightarrow 6p_{3/2},f=4$,   we pump the atoms into a 50/50 mix of $\ket{4,4}$ and $\ket{4,-4}$. Using a modulator to rapidly switch the probe frequency seen by the atoms between $\nu_{\textrm c}+\delta\nu_+$ and $\nu_{\textrm c}+\delta\nu_-$ we can acquire data from both transitions, and average to find the unshifted 
clock frequency. In this approach the atom number for each transition is reduced by a factor of two so $\sigma$ in eq. (\ref{eq.sigma}) is increased by $\sqrt2$. If the alternation is implemented at frequency $f_{\textrm s}\sim 10~\textrm{kHz}$ this will introduce intermodulation noise at $2 f_{\textrm s}$\cite{Audoin1991}. This has has been accounted for in Sec.~\ref{sec.stability}.

Alternatively we may probe two different atomic samples, either in nearby regions of one vacuum cell, or in separate vacuum cells, each on one of the two stretched state transitions. In this case 
there is a potential sensitivity to the difference in the magnetic field at each location. If the magnetic field variation is $ \pm\delta B/2$ the average of the measured frequencies will be 
$$
\nu_{\textrm c}+\delta\nu= \nu_{\textrm c}+ \frac{\mu_B}{2\pi\hbar}\delta B.
$$
The sensitivity coefficient is $\beta=1400 ~\textrm{Hz}/(10^{-7}~\textrm{T})$. This effect can be suppressed by alternating back and forth between which region probes which transition and again taking the average. However, this will not cancel the effect of time varying magnetic field gradients.

\subsection{DC electric field}

The static scalar polarizability of the Cs 6s ground state has been accurately measured 
to be\cite{Amini2003}
$\alpha_0^{(6s)}=59.4~\textrm \AA^3.$ The tensor polarizability of the ground state is negligible for the following analysis. 
Accounting for scalar and tensor contributions we calculate the static polarizability of the $5d_{5/2}\ket{6,6}$ state to be 
$\alpha_{5d}=30.8~\textrm \AA^3$ The sensitivity coefficient for the clock transition is 
$\beta= ~4.7~ \textrm{Hz/(V/m)}.$

\subsection{Lattice wavelength}
\label{sec.latwavestab}

As is shown in Fig.~\ref{fig.dynamicStark2}
the differential polarizability of the excited and ground states is approximately 
$1.4\times 10^{-4}~\textrm{\AA}^3$
per MHz of  lattice laser shift from the magic point. This translates into a sensitivity coefficient $\beta=0.0078~\textrm{Hz}/\textrm{MHz}$ assuming the atoms have a temperature of $1~\mu\textrm{K}$.

\subsection{Probe beam power}

\begin{figure}[!t]
    \centering
    \includegraphics[width=0.5\textwidth]{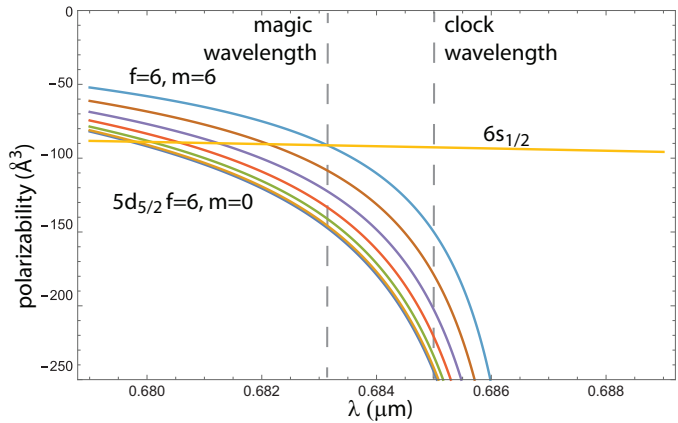}
    \caption{Polarizability of ground and excited states near the clock wavelength.  }
  \label{fig.magic685}
\end{figure}

The clock frequency is sensitive to the probe beam power due to differential Stark shifts of the ground and excited states. This is a relatively small effect since the clock wavelength is close to a magic wavelength as can be seen in Fig.~\ref{fig.magic685}.
The differential polarizability at the clock wavelength is $\delta\alpha=-57.5~\textrm \AA^3$. For a probe beam intensity that provides unit saturation we find  a sensitivity coefficient per \% change of the probe power of $\beta=0.11~\textrm{Hz}/\%$.\\

\subsection{Blackbody radiation}

Accurate calculations of dynamic Stark shifts due to blackbody radiation can be found in \cite{Farley1981}. At $T=300~\textrm K$
the shifts of the $6s_{1/2}$ and $5d_{5/2}$ states are -3.589 and 5.315 Hz, respectively. The differential shift is thus $\delta\nu=8.904 ~\textrm{Hz}$. Since the blackbody shift scales as $T^4$ the sensitivity at 300 K is $\beta =0.12~\textrm{Hz}/\textrm{K}$.

\section{Discussion}

We have introduced a cesium lattice optical clock using a forbidden quadrupole transition. System complexity is considerably reduced compared to a state of the art atomic optical clock using alkaline earth atoms. Predicted performance is intermediate between microwave clocks and existing atomic optical clocks. The use of a high atom number, combined with magic optical trapping, provides a path to high stability that exceeds that of Cs fountain clocks at short times\cite{Heavner2014}, as well as competitive long term stability, all in a more compact package. 

The proposed cesium clock requires three lasers at wavelengths of 685, 803, and 852 nm, all of which are available from semiconductor sources. Acousto-optic and electro-optic modulators can be used to provide the frequency shifts needed for cooling, optical pumping, and repumping operations.  Cesium can be conveniently laser cooled directly from a room temperature source in a compact vacuum package. We envision implementing the lattice build-up resonator described in Sec.~\ref{sec.traparray} in a compact monolithic form factor using a fiber Fabry-Perot resonator. To complete the system, an optical frequency comb is referenced to the 685 nm probe laser. Small form factor, yet high performance architectures for frequency combs\cite{Sinclair2015} together with progress in high stability optical oscillators without the need for bulky and vibration sensitive cavity stabilization\cite{WLoh2020} point towards the potential for a complete clock system in a compact form factor.  Progress in engineering of state of the art Sr clocks has also resulted in high performance in a reduced form factor\cite{Ohmae2021,FGuo2021}. Compared to the CLOC,
Sr requires five laser wavelengths (and often multiple lasers per wavelength), as well as a more complex atomic source based on an oven and a Zeeman slower. The narrow width of the Sr clock transition, while advantageous for reaching the highest possible performance, also implies the need for a more stable probe laser, which in turn requires relatively bulky and vibration sensitive cavity stabilization. 

Rubidium has a similar atomic structure to Cs and may in principle be used in an analogous fashion. The corresponding clock transition in $^{87}$Rb is  $5s_{1/2},f=2\rightarrow 4d_{5/2},f=4$ at 517 nm
with cycling decay via $5p_{3/2},f=3$ back to the ground state. Analysis shows that there are several magic wavelengths for both dark and bright optical traps. However, the atomic parameters are not as favorable as they are in Cs.  The lifetime of the $4d_{5/2}$ level is 89 ns, corresponding to a linewidth of 1.8 MHz, which is more than an order of  magnitude broader than the upper clock state in Cs. This is partially compensated for by the increase in $\dot N$ in eq. (\ref{eq.sigma}). Assuming the same number of atoms and detection efficiency as for Cs we find $\sigma(\tau)=2.4\times 10^{-15}/\sqrt\tau$ which is a factor of 2.9 larger than for Cs. While still attractive for a clock implementation the requirement of resolved sideband excitation to suppress Doppler broadening would imply more than an order of magnitude higher trap vibrational frequencies. Since the optical trap frequencies scale with the square root of the optical intensity times the polarizability, a trap intensity more than two orders of magnitude higher would be needed, which would be difficult to implement. 

Finally we note that the atomic sample  with individually trapped Cs atoms arranged at $\mu\textrm m$ scale spacings is closely related to systems for quantum simulation and computation that use Rydberg interactions to generate entanglement. Recent advances have demonstrated multi-particle, highly  entangled states with both analog\cite{Ebert2015,Zeiher2015,Omran2019} and digital control modalities\cite{Graham2021}. Combining the architecture described here with Rydberg mediated entanglement may provide a feasible route towards  quantum enhanced clock performance.\\  

\section{Acknowledgements}
We gratefully acknowledge Andrew Jayich for a critical reading of the manuscript. This material is based upon work supported by the Air Force Office of Scientific Research under award number FA9550-21-1-0034.

\bibliography{atomic,saffman_refs,rydberg,qc_refs,optics}

\end{document}